# Multiple Routes for Glutamate Receptor Trafficking: Surface Diffusion and Membrane Traffic Cooperate to Bring Receptors to Synapses

Laurent Cognet[1], Laurent Groc[2], Brahim Lounis[1] and Daniel Choquet[2]

[1]*Centre de Physique Moléculaire Optique et Hertzienne, CNRS (UMR 5798) and Université Bordeaux 1, 351 Cours de la Libération, 33405 Talence Cedex, France*
[2]*Physiologie Cellulaire de la Synapse, CNRS (UMR 5091) and Université Bordeaux 2 Institut François Magendie, 33077 Bordeaux Cedex, France*
*Corresponding author. E-mail, dchoquet@u-bordeaux2.fr

**Abstract**

Trafficking of glutamate receptors into and out of synapses is critically involved in the plasticity of excitatory synaptic transmission. Endocytosis and exocytosis of receptors have initially been thought to account alone for this trafficking. However, membrane proteins also traffic through surface lateral diffusion in the plasma membrane. We describe developments in electrophysiological and optical approaches that have allowed for the real time measurement of glutamate receptor surface trafficking in live neurons. These include (i) specific imaging of surface receptors using a pH sensitive fluorescent protein, (ii) design of a photoactivable drug to inactivate locally surface receptors and monitor electrophysiologically their recovery, and (iii) application of single molecule fluorescence microscopy to directly track the movement of individual surface receptors with nanometer resolution inside and outside synapses. Altogether, these approaches have demonstrated that glutamate receptors diffuse at high rates in the neuronal membrane and suggest a key role for surface diffusion in the regulation of receptor numbers at synapses.

AMPA (α-amino-3-hydroxy-5-methyl-4-isoxazole propionate) and NMDA (*N*-methyl-D-aspartate) ionotropic glutamate receptors (AMPARs, NMDARs, abbreviated as a class as iGluRs) are ligand-activated cation channels concentrated in the postsynaptic density (PSD), which mediate fast excitatory neurotransmission in the central nervous system (*1*). Concentration of these receptors at PSDs is thought to result from their stabilization by interactions with specific intracellular scaffolding proteins and cytoskeletal elements (*2, 3*). In fact, both AMPARs and NMDARs constitutively cycle in and out of the postsynaptic membrane at rapid, albeit different, rates through processes of endo- and exocytosis. Moreover, the turnover rate of the whole cycle is highly regulated by neuronal activity, phosphorylation, and scaffold and accessory proteins [reviewed in (*4, 5*)]. As a result, regulation of the balance between endo- and exocytosis, especially for AMPARs, has initially been thought to account alone for the rapid variations in receptor composition of the PSD during synaptic plasticity. However, the observation that AMPARs could be highly mobile in the neuronal plasma membrane (*6*) together with the demonstration that endocytosis (*7-9*) and exocytosis (*10*) likely occurs outside synapses suggest that lateral diffusion could act as a complementary trafficking pathway for the regulation of iGluRs numbers at synapses (*11, 12*).

In this perspective, we describe the recently developed electrophysiological and optical approaches that have allowed measururement of surface receptor trafficking inside and outside synapses, and permitted differentiation of diffusion-based trafficking from from endo- and exocytotic processes.

Electrophysiological approaches are classically used to measure the participation of receptors to synaptic transmission. A number of refinements, based on genetic or pharmacological control of a receptor's functional properties, were developed to measure the trafficking dynamics of functional receptors. Specific mutations permit the electrophysiological tagging of recombinant AMPARs that differ in their rectification properties from endogenous receptors. Their incorporation into the postsynaptic membrane can thus be electrophysiologically monitored in brain slices or in vivo (*13, 14*) when synapses were subjected to various activation patterns. This approach cannot however measure basal receptor turnover and relies on exogenously expressed receptors.

Methods to electrophysiologically study native receptor trafficking were also developed. They rely on the recordings of functional receptor recovery rates after irreversible block by pharmacological reagents. Although a disadvantage of these methods is the need to use receptor antagonists, they provide unique information on trafficking of functional receptors. The irreversible NMDAR blocker MK-801 was used to block either all cell surface or only synaptic NMDARs. This allowed the deduction that intracellular NMDAR insertion at the cell plasma membrane does not occur within a 30-minute period. More importantly, the fast recovery of synaptic NMDAR currents suggested that extrasynaptic NMDARs moved into synapses through lateral diffusion (*15*).

Adesnik *et al.* (*16*) used a membrane-impermeable and photoreactive AMPAR antagonist (6-azido-7-nitro-1,4-dihydroquinoxaline-2,3-dione, ANQX). IGlobal or focal photoactivation by ultraviolet (UV) light allows measurement the recovery of currents after specific inactivation of either all or a subpopulation of surface AMPARs. The turnover rate of synaptic receptors with intracellular pools was found to be surprisingly slow, in the time scale of several hours, differing from previous results. In contrast, AMPARs were shown to be rapidly exocytosed to the plasma membrane in the somatic compartment. Somatic surface AMPARs also diffused laterally at high rate. Thus, it is possible that new functional AMPARs in the dendrites come from receptors that are exocytosed at the soma and travel down dendrites by lateral diffusion. It will be interesting to determine whether exchange of receptors between synaptic and neighbouring extrasynaptic membrane can be observed with this technique. It is also noteworthy to mention that local AMPARs synthesis in dendrites and insertion into synaptic plasma membranes has been observed (*17*).

Along with biochemical approaches, these experiments have demonstrated that receptor recycling and lateral diffusion are both important contributors to receptor trafficking. Both biochemical methods and electrophysiological methods have limitations. Biochemical approaches lack spatial and temporal resolution, whereas electrophysiological ones only detect cell surface expressed functional receptors without giving a direct picture of receptors spatial organization.

Optical methods have been used to track iGluRs with high resolution in different subcellular compartments. For this purpose, receptors have been tagged either with green fluorsecent protein (GFP) variants, an alpha-bungarotoxin binding site (*18*), a tetracystein tag that binds bi-arsenical dyes (*17*), a biotine ligase site (*19*), or receptor-specific antibodies conjugated to fluorophores. The development of GFP-tagged receptors had initially raised great hopes for real-time optical visualization of receptor trafficking. For instance, in immature neurons, GFP-tagged NMDARs and AMPARs subunits were detected in mobile transport packets recruited rapidly and independently to nascent synapses (*20, 21*). In mature synapses, GFP-tagged GluR1, a subunit of AMPAR, relocated to spines during protocols of that produce synaptic potentiation (*22*). However, imaging of GFP-tagged receptors bears serious limitations. First, these exogeneously expressed receptors might have different trafficking properties than do endogenous ones. Second, sparsely distributed GFP-tagged receptors are difficult to detect with standard microscopes (*23*). Third, a large fraction of iGluRs, and this is especially true for AMPARs, are located in intracellular compartments and surface GFP-tagged receptors cannot be easily differentiated from internal ones. This latter drawback has been partly overcome by the use of a pH-sensitive form of GFP (pHLuorin). This form of GFP can be used to discriminate surface from internal receptors because of the differential fluorescence due to the acidic pH in various intracellular compartments (for example, endoplasmic reticulum, Golgi, and secretory vesicles) and the almost neutral pH in the extracellular environment. Experiments with pHluorin-tagged AMPAR have shown that receptor endocytosis occurs at extrasynaptic sites (*9*).

The dynamics of receptor movements in different subcellular compartments can be studied with receptors tagged by GFP variants using fluorescence recovery after photobleaching (FRAP) or photoactivation approaches, as recently exemplified for dopamine receptors (*24*), γ-amino-butyric acid receptors (GABARs) (*25*), and iGluRs in *Drosophila* (*26*). These studies have also revealed an important role for receptor lateral diffusion in the regulation of receptor numbers at synapses. However, such approaches can only provide information on the average behaviors of rather large receptors assemblies with diffraction-limited spatial resolutions (few hundreds of nanometers).

In contrast, single particle and single molecule detection (SMD) techniques allow direct measurement of the movements of individual surface AMPARs (*6, 27*), NMDARs (*28*), metabotropic glutamate receptors (mGluRs) (*29*), as well as non-glutamatergic receptors (*30*) with nanometer resolution. Movements of receptors at the neuronal surface have initially been recorded by tracking submicron sized latex particles manipulated with optical tweezers (*6, 31*). The size of the particles however precluded access to synaptic receptors. This drawback has been partly overcome by replacing the bead with individual nanometer-sized fluorescent objects, such as an organic fluorophore (*27*) or a semiconductor nanocrystal, also known as "quantum dots" (*28, 30*). Although organic dyes are very small (typically 1 nm), the total amount of photons they can emit is limited, restricting the recording times to typically a few seconds. On the contrary, although functionalized quantum dots are bulkier (~20 to 30 nm) and display a characteristic blinking behavior, their strong photoresistance allows for minute long recordings.

In SMD experiments, endogenous or transfected receptor subunits are labeled at dilute concentrations with individual fluorophores through a ligand (typically an antibody) that binds to the extracellular domain of a native receptor (*27*) or to a genetically engineered tag (*19*). Tracking the movement of individual (or small assemblies of) receptors in different submembrane compartments is thus achieved. Although SMD has so far been applied predominantly to to study cell surface receptor movements, SMD can also be applied to follow intracellular movements.

SMD provides an important advantage over techniques that track large receptor ensembles , namely the possibility of tracking molecules with a positional accuracy only limited by the signal-to-noise ratio at which the molecules are detected, in practice below 40 nm in live cells. This allows endogenous iGluR trajectories to be measured inside synapses (*27, 28*), even if synapse diameters are ~300 nm. Moreover, by removing the averaging inherent to ensemble measurements, SMD yields a measure of the full distributions of iGluR mobilities inside and outside synapses. Surface iGluR diffusion rates span several orders of magnitude, revealing the existence of different subpopulations with distinct diffusion behaviors. Interestingly, iGluR subtypes display different mobilities as NMDARs are less mobile than AMPARs (*28*) or mGluRs (XX citation). However, for all receptor types, both immobile and mobile receptors are found in synaptic and extrasynaptic regions, where they display specific diffusion characteristics. On the one hand, most extrasynaptic AMPARs exhibit free Brownian diffusion (*27*). They can explore large surface areas of the neuronal membrane and travel along several micrometers. This finding was corroborated by the eletrophysiological measurements described above (*16*). On the other hand, in synapses, rapidly mobile iGluRs are also found, but with a distinctive diffusion signature: Their mobility is restricted to areas equivalent to or smaller than that of the synapse (*27*). These specific diffusion properties observed within structures that cannot be resolved by an optical microscope could only be studied by SMD where any ensemble measurements would have failed.

SMD also revealed that receptors exchange rapidly by lateral diffusion between submicron neighboring compartments, such as the synaptic, peri- and extrasynaptic spaces (*27, 30*). In other words, lateral diffusion allows receptors to be locally and rapidly added or removed from their functional site at synapses (*12*). It remains to be further determined whether mobile and immobile receptors bear similar functional properties.

Receptor lateral diffusion is likely to play an important role in the physiology of synaptic transmission. Indeed, receptor mobility is highly regulated by a number of pharmacological treatments known to modify synaptic transmission (*27, 28*) (that is elevated calcium, depolarization, neuronal activity, and activation of signaling cascades). However, a direct link has up to now not been established.

SMD techniques still suffer from a number of shortcomings. Organic dyes have poor photostabilities and the bulky size of functionalized quantum dots might sterically hinder receptor diffusion in highly confined domains, such as the synaptic cleft. Furthermore, attachment of these labels to native receptors still usually relies on the availability of antibodies, which are rather bulky ligands, targeted to extracellular domains. An additional difficulty arises from the inability to unambiguously distinguish surface versus intracellularly located labels. These limitations have restricted the application of SMD studies to cultured cells. Although luminescence properties of individual quantum dots should allow for their visualization in tissue, such as brain slices or in vivo, their size would likely limit their penetration, diffusion, or both. The future developments of SMD will thus require smaller and more photostable functionalized biocompatible labels, as well as smaller ligands. For the former, a possible route is provided by metallic nanoparticles, which have been recently detected by purely optical means at sizes as small as 1.4 nm in diamter in polymers (*32, 33*) and 5 nm in diameter in live cells (*34, 35*). For the latter, new ligands such as toxins (*18*) or recombinant antibody Fab fragments are future directions.

In conclusion, a combination of state-of-the-art optical and electrophysiological approaches have allowed the real-time measurement of receptor trafficking in live neurons. These have shed light on the key role played by receptors lateral diffusion at the neuronal surface. This process likely acts as a complementary pathway to membrane exchange events involved in the regulation of receptor numbers at synapses. In the future, further technical developments requiring the combination of chemistry, physics, and biology will be needed to improve our ability to measure the movements of individual receptors under physiological conditions.

Fig. 1. Schematic representation of AMPAR trafficking in neurons and experimental approaches used to track receptors in real time. Three main pathways of AMPAR trafficking into and out of synapses have been identified, endocytosis, exocytosis, and lateral diffusion. Surface AMPARs

can diffuse laterally at a high rate and explore vast dendritic areas by random movements. At smaller scales, they can also move within the synapses (upper inset) and exchange between the extrasynaptic and the postsynaptic membrane. Immobile AMPARs are likely stabilized by scaffold elements. AMPARs also cycle between intracellular vesicles and surface membrane. AMPARs are endocytosed in the vicinity of the synapse and in the extrasynaptic membrane. The sites of AMPARs insertion whether solely at the soma, at dendrites, or at the synapse is still debated. (Lower inset) Receptor trafficking can be measured in real time by three main approaches. First, genetically engineered receptors can be tagged and monitored electrophysiologically (for example, by differences in the tagged receptors' rectification properties) or imaged with fluorescent reporters (for example green fluorescent protein, GFP). Second, pharmacological tools can selectively block surface receptors. For instance, a photoactivable drug may locally inactivate surface receptors, providing a mechanism for investigating the trafficking of functional receptors. Third, single molecule fluorescence microscopy can track the movement of individual (or a very small group) of surface receptors with nanometer resolution. In this case, surface receptors are detected by specific antibodies (directed against an extracellular epitope), which are labelled either with an organic fluorophore (for example, cyanine dye) or a nanometer-size particle (for example, a quantum dot).

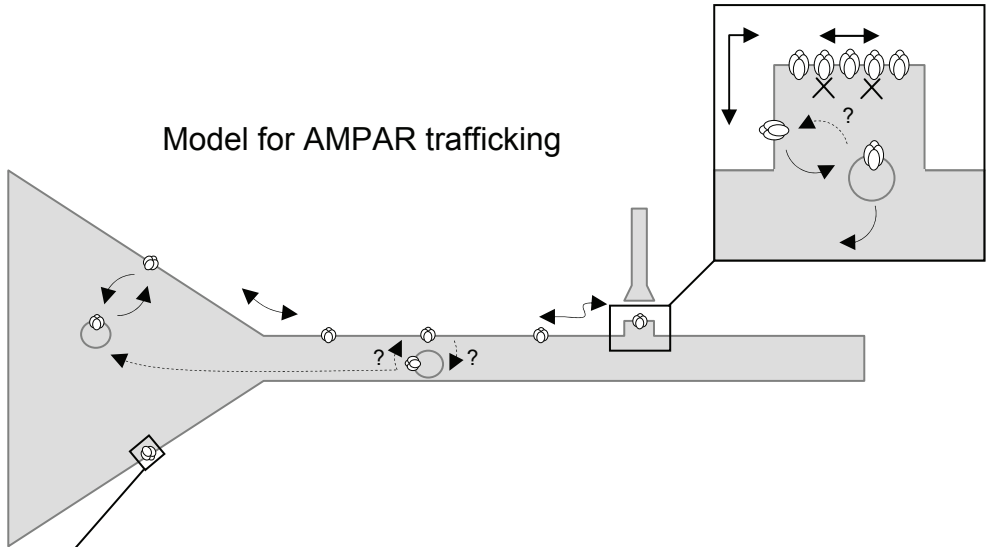

Model for AMPAR trafficking

How to detect receptor trafficking in real time?

Genetically engineered receptors

Electrophysiol. tags
e.g. rectification change

Fluorescent tags
e.g. GFP, pHluorin

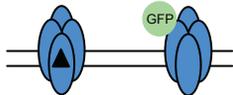

Pharmacological tools
(e.g. photoactivable blocker)

UV light

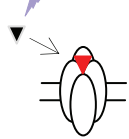

Single molecule / QD tracking

Quantum dot

e.g. cyanine dye

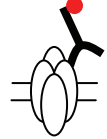 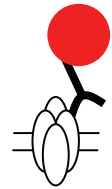